\documentclass[aps,prb,reprint,superscriptaddress,longbibliography]{revtex4-2}

\usepackage{graphicx}
\usepackage{amsmath,amssymb,bm}
\usepackage{dcolumn}
\usepackage{hyperref}
\usepackage{xcolor}
\usepackage{braket}
\usepackage{balance}

\begin{document}

\title{Observation of complex orbital order in MnV$_2$O$_4$}

\author{Chihaya Koyama}
\email{koyama-chihaya@g.ecc.u-tokyo.ac.jp}
\affiliation{Department of Advanced Materials Science, The University of Tokyo, Kashiwa 277-8561, Japan.}

\author{Shunsuke Kitou}
\email{kitou@edu.k.u-tokyo.ac.jp}
\affiliation{Department of Advanced Materials Science, The University of Tokyo, Kashiwa 277-8561, Japan.}

\author{Taishun Manjo}
\affiliation{Japan Synchrotron Radiation Research Institute (JASRI), SPring-8, Hyogo 679-5198, Japan.}

\author{Yuiga Nakamura}
\affiliation{Japan Synchrotron Radiation Research Institute (JASRI), SPring-8, Hyogo 679-5198, Japan.}

\author{Takeshi Hara}
\affiliation{Department of Physics, Tohoku University, Sendai 980-8578, Japan.}

\author{Naoyuki Katayama}
\affiliation{Department of Physics, Okayama University, Okayama 700-8530, Japan.}

\author{Takuro Katsufuji}
\affiliation{Department of Physics, Waseda University, Tokyo 169-8555, Japan.}
\affiliation{Kagami Memorial Research Institute for Materials Science and Technology, Waseda University, Tokyo 169-0051, Japan.}

\author{Ryotaro Arita}
\affiliation{Department of Physics, The University of Tokyo, Tokyo 153-8904, Japan.}
\affiliation{RIKEN Center for Emergent Matter Science (CEMS), Wako 351-0198, Japan.}

\author{Yusuke Nomura}
\affiliation{Institute for Materials Research (IMR), Tohoku University, Sendai 980-8577, Japan.}
\affiliation{International Quantum Materials Center (Q²),Advanced Institute for Materials Research (WPI-AIMR), Tohoku University, Sendai 980-8577, Japan.}

\author{Hiroshi Sawa}
\affiliation{Nagoya Industrial Science Research Institute, Nagoya 460-0008, Japan.}

\author{Taka-hisa Arima}
\affiliation{Department of Advanced Materials Science, The University of Tokyo, Kashiwa 277-8561, Japan.}
\affiliation{RIKEN Center for Emergent Matter Science (CEMS), Wako 351-0198, Japan.}

\begin{abstract}
Orbital ordering in vanadium spinel oxides with a geometrically frustrated pyrochlore structure has been a subject of controversy, owing to competing theoretical models and the absence of direct experimental evidence. Here we combine high-precision single-crystal synchrotron x-ray diffraction with core differential fourier synthesis to visualize the valence electron density (VED) of the orbital-ordered ground state in real space. By carefully investigating multiple-scattering artifacts, we identify the low-temperature structure as belonging to the $I4_1/amd$ space group. The reconstructed VED around the V sites reveals an orbital-ordered state distinct from both previously proposed real- and complex-orbital models. Our results resolve the long-standing controversy in MnV$_2$O$_4$ and establish a route to identifying orbital states in frustrated spin-orbital systems.
\end{abstract}

\maketitle

\begin{figure}[t]
\centering
\includegraphics[width=\columnwidth]{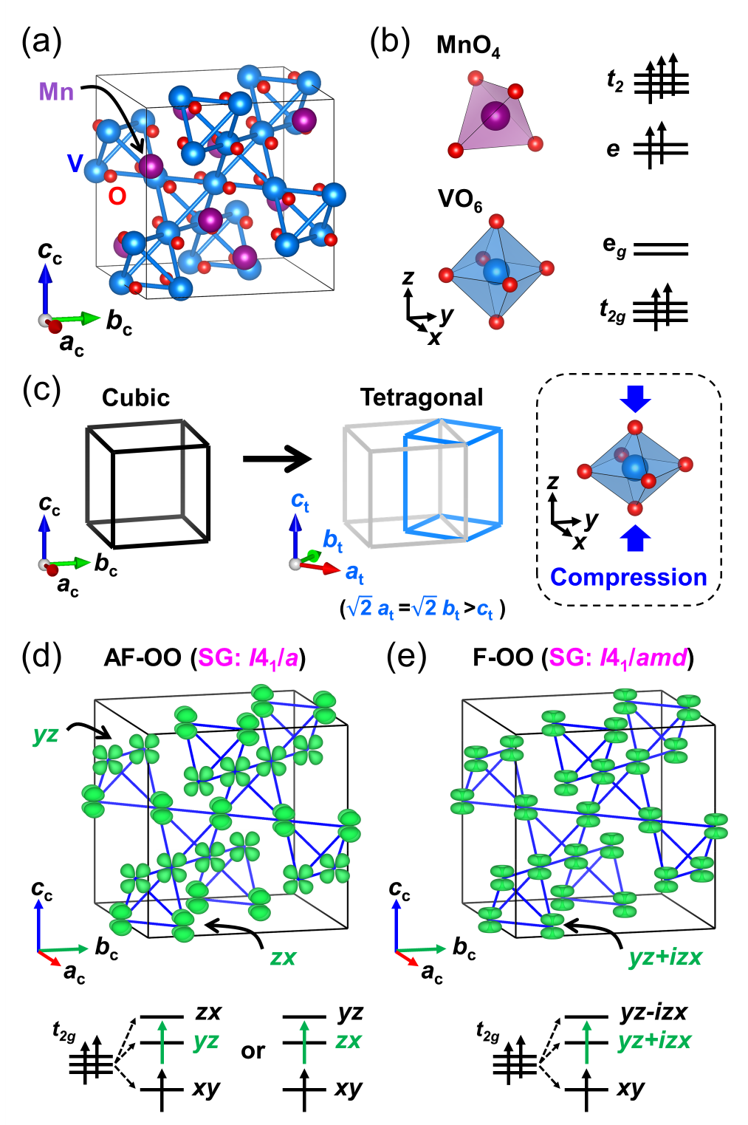}
\caption{(a) Crystal structure of MnV$_2$O$_4$ in the high-temperature cubic phase. (b) Schematic illustration of the orbital states of an Mn$^{2+}$ ion in an MnO$_4$ tetrahedron and a V$^{3+}$ ion in a VO$_6$ octahedron. Here, we define the quantization axes as $x \parallel a_c$, $y \parallel b_c$, and $z \parallel c_c$. (c) Schematic illustration of the lattice distortion associated with the structural phase transition, where the subscripts c and t in the unit cell axes denote the cubic and tetragonal phases, respectively. In the tetragonal phase, the VO$_6$ octahedra are compressed along the $z$ axis. (d),(e) Two theoretical models for orbital ordering of MnV$_2$O$_4$ in the ground state, corresponding to antiferro-orbital-ordering (AF-OO) and ferro-orbital-ordering (F-OO), with space groups (SG) $I4_1/a$ and $I4_1/amd$, as proposed in Refs.~\onlinecite{Tsunetsugu2004,Tchernyshyov2004}, respectively. Here, assuming that the energy splitting caused by the tetragonal distortion is dominant, the $d_{xy}$ orbital is lowered, and only the other $t_{2g}$ electron responsible for the orbital ordering, indicated by the green arrows, is shown schematically.}
\label{fig:fig1}
\end{figure}

When $d$ electrons with orbital degeneracy are placed in a three-dimensional geometrically frustrated network of corner-sharing tetrahedra, referred to as the pyrochlore structure, the interplay between spin and orbital gives rise to nontrivial ground states~\cite{Khomskii2003}. A typical example is the family of Mott-insulating vanadium spinel oxides $A$V$_2$O$_4$ ($A$ is a divalent cation), which have attracted considerable attention and have been extensively studied. In $A$V$_2$O$_4$, V$^{3+}$ ions with a $3d^2$ electronic configuration form a pyrochlore structure [Fig.~\ref{fig:fig1}(a)]. When nonmagnetic ions such as Mg, Zn, and Cd occupy the $A$ site, the V$^{3+}$ $S$=1 spins exhibit antiferromagnetic order~\cite{Wheeler2010,Niitaka2013,Maitra2007,Lee2004,Kiswandhi2014,Tsunetsugu2003,Tchernyshyov2004}. In contrast, when magnetic ions such as Fe$^{2+}$ and Mn$^{2+}$ occupy the $A$ site, noncollinear ferrimagnetism shows up~\cite{Suzuki2007,Katsufuji2008,Garlea2008,Nii2012,MacDougall2012,Okabayashi2015,Nakano2025}. In MnV$_2$O$_4$, Mn$^{2+}$ ($S=5/2$) ions at tetrahedral sites adopt a high-spin 3$d$$^{5}$ configuration without orbital degree of freedom (ODF), whereas V$^{3+}$ ($S=1$) ions at octahedral sites retain active ODF in the $t_{2g}$ orbitals [Fig.~\ref{fig:fig1}(b)]. At $T_{\mathrm{OO}}=53$ K, the system shows a cubic-to-tetragonal structural transition, resulting in a tetragonal phase compressed along the $c$ axis, accompanied by the formation of orbital order (OO) at the V sites [Fig.~\ref{fig:fig1}(c)]~\cite{Suzuki2007,Nii2012}. However, the local orbital states of the V$^{3+}$ ions and their ordering pattern have remained unresolved for many years~\cite{Suzuki2007,Zhou2007,Garlea2008,Sarkar2009,Wei2010,Nii2012,Wei2015,Dey2016,Dey2017,Jo2017,Matsuura2017, Matsuura2018}.

As the ground state of MnV$_2$O$_4$, several OO models have been proposed. Tunetsugu and Motome proposed an antiferroic orbital ordering (AF-OO) model [Fig.~\ref{fig:fig1}(d)]~\cite{Tsunetsugu2003, Tsunetsugu2004}, where $d_{yz}$ and $d_{zx}$ orbitals are alternately ordered along the one-dimensional chains running along the $\langle 110\rangle$ directions as a result of the interplay between $dd$ superexchange interactions and geometric frustration. In this model, the $d$-glide symmetry perpendicular to the $\langle 110\rangle$ direction is broken, leading to the $I4_1/a$ space group. The observation of reflections disagreeing the $d$-glide symmetry in diffraction experiments suggests that AF-OO is stabilized in the ground state of MnV$_2$O$_4$~\cite{Suzuki2007,Nii2012}. However, inelastic neutron scattering experiments suggests that this simple AF-OO picture is insufficient, because the interchain exchange inferred from spin-wave analysis is too large to be reconciled with the weakly coupled one-dimensional chains assumed in the AF-OO model~\cite{Chung2008}. On the other hand, Tchernyshyov proposed a ferroic orbital ordering (F-OO) model [Fig.~\ref{fig:fig1}(e)]~\cite{Tchernyshyov2004}, where spin-orbit coupling plays a major role, and one of the two V $t_{2g}$ electrons occupies a complex orbital  ($d_{yz}\pm i d_{zx}$). This model satisfies the $I4_1/amd$ space-group symmetry and the V site possesses orbital angular momentum along the $c$ axis. It is also qualitatively consistent with x-ray magnetic circular dichroism (XMCD) measurements that revealed an orbital magnetic moment of $0.15\,\mu_\mathrm{B}$ per V ion~\cite{Okabayashi2015}.

Although each of these models is consistent with part of the experimental observations, a unified interpretation of the OO ground state has yet to be established. The central difficulty is that crystallographic symmetry and magnetic response constrain the possible orbital configurations, but do not uniquely determine the occupied orbital wave function.  A direct real-space probe of valence-electron density (VED) is therefore essential for resolving the OO problem in MnV$_2$O$_4$. In this study, we performed single-crystal synchrotron x-ray diffraction (XRD) to determine the low-temperature space group of MnV$_2$O$_4$ and directly visualized the orbital-ordered state through VED analysis.

A single crystal of MnV$_2$O$_4$ was grown by the floating-zone method~\cite{Gleason2014}. XRD experiments were conducted at the BL02B1 beamline~\cite{Sugimoto2010} of the SPring-8 synchrotron facility in Japan. A single crystal used in the experiments was $40 \times 40 \times 40\,\mu\mathrm{m}^3$. The x-ray wavelength was 0.3095 \AA, and temperature variations were controlled using N$_2$/He gas blowing. Diffraction patterns were recorded using a two-dimensional CdTe PILATUS detector with a dynamic range of $\sim 10^6$. The intensities of Bragg reflections were collected by CrysAlisPro~\cite{CrysAlisPro}. Intensities of equivalent reflections were averaged using SORTAV~\cite{Blessing1989}, and structural parameters were refined by Jana2006~\cite{Petricek2014}. Here, by utilizing only high-angle reflections ($\sin\theta/\lambda > 0.6\,\mathrm{\AA}^{-1}$), where the contribution of spatially spread valence electrons to XRD is negligible, structural parameters including atomic displacement parameters were obtained with high accuracy.
The core differential fourier synthesis (CDFS) method~\cite{Kitou2017,Kitou2020} was used to extract the VED distribution around each atomic site. The [Ar]-type electrons of the Mn and V atoms, as well as the [He]-type electrons of the O atoms, were regarded as core electrons. In Jana2006, three basic Slater-type orbital (STO) libraries are supplied. In this study, we used the STO-COPPENS library for the CDFS analysis. The crystal structure and VED distributions were visualized using VESTA~\cite{Momma2011}.

\begin{figure}[t]
\centering
\includegraphics[width=\columnwidth]{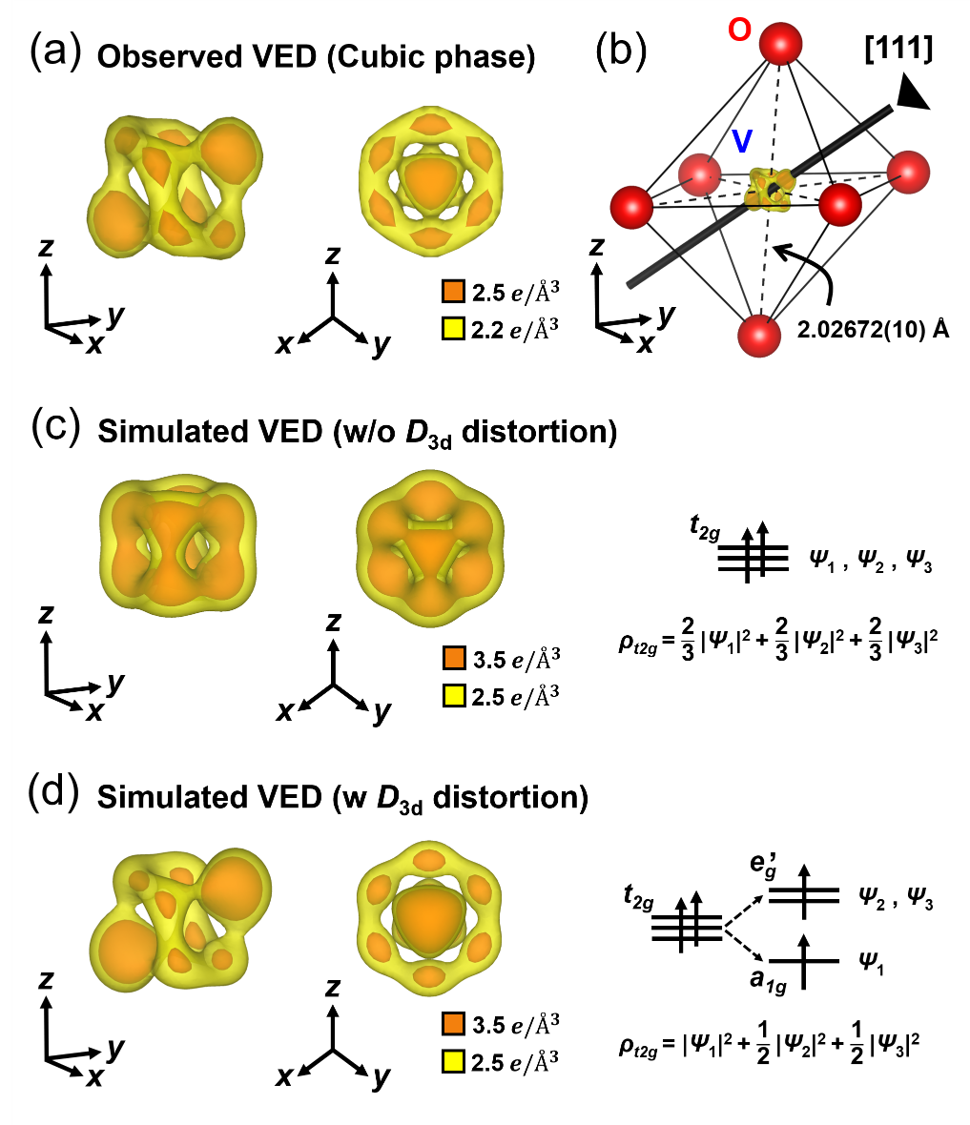}
\caption{(a) Observed VED distribution around V at 160 K in the cubic phase. Yellow and orange isodensity surfaces represent low and high electron-density levels, respectively.
(b) A VO$_6$ octahedron showing the VED around the V site at 160 K cubic phase. The VO$_6$ octahedron is elongated in the [111] direction. (c) Simulated VED distribution expected for threefold-degenerate $t_{2g}$ orbitals. (d) Simulated VED distribution assuming $D_{3d}$ distortion. A schematic diagram of the orbital energy levels is also shown. The pronounced anisotropy in the simulated VED indicates that the trigonal distortion lifts the $t_{2g}$ degeneracy into an $a_{1g}$ singlet and an $e'_g$ doublet, leaving the $e'_g$ doublet as the active ODF.}
\label{fig:fig2}
\end{figure}

We first examine the VED distribution at 160 K in the high-temperature cubic phase as a reference for understanding the orbital state of the system. The VED distributions around  O and Mn sites are nearly isotropic, consistent with the $2s^2 2p^6$ valence electron configuration and the orbitally inactive high-spin Mn$^{2+}$ ($3d^5$) states, respectively (see Supplemental Information~\cite{sup}). In contrast, a pronounced anisotropic VED distribution is observed around the V site [Fig.~\ref{fig:fig2}(a)]. Here, we take the V site at the internal coordinate $(1/2, 1/2, 1/2)$ as a representative site. The $x$, $y$, and $z$ axes define the local coordinate system, with $x\parallel a_{\mathrm{c}}$, $y\parallel b_{\mathrm{c}}$, and $z\parallel c_{\mathrm{c}}$, as shown in Fig.~\ref{fig:fig1}(a). Structural analysis reveals that the VO$_6$ octahedra are slightly elongated along the threefold axis [Fig.~\ref{fig:fig2}(b)]. Although this distortion lifts the threefold degeneracy of the $t_{2g}$ orbitals into $a_{1g}$ and $e'_g$ states from the symmetry viewpoint, we first calculate the VED distributions assuming threefold-degenerate $t_{2g}$ orbitals, \textit{i.e.}, neglecting the trigonal distortion [Fig.~\ref{fig:fig2}(c)]. The simulated VED fails to reproduce the observed one [Fig.~\ref{fig:fig2}(a)]. In contrast, the simulated VED assuming the trigonal distortion in Fig.~\ref{fig:fig2}(d) agrees well with the observed VED around the V site, which suggests that one electron occupies the lower-lying $a_{1g}$ orbital, leaving the remaining ODF in the higher-lying $e'_g$ doublet [Fig.~\ref{fig:fig2}(d)]. This VED resembles that observed in the cubic phase of the related compound FeV$_2$O$_4$~\cite{Manjo, Koyama}. The remaining questions are what lifts the degeneracy of the $e'_g$ doublet at low temperatures and how the resulting orbital states are spatially arranged to establish OO.

\begin{figure}[t]
\centering
\includegraphics[width=\columnwidth]{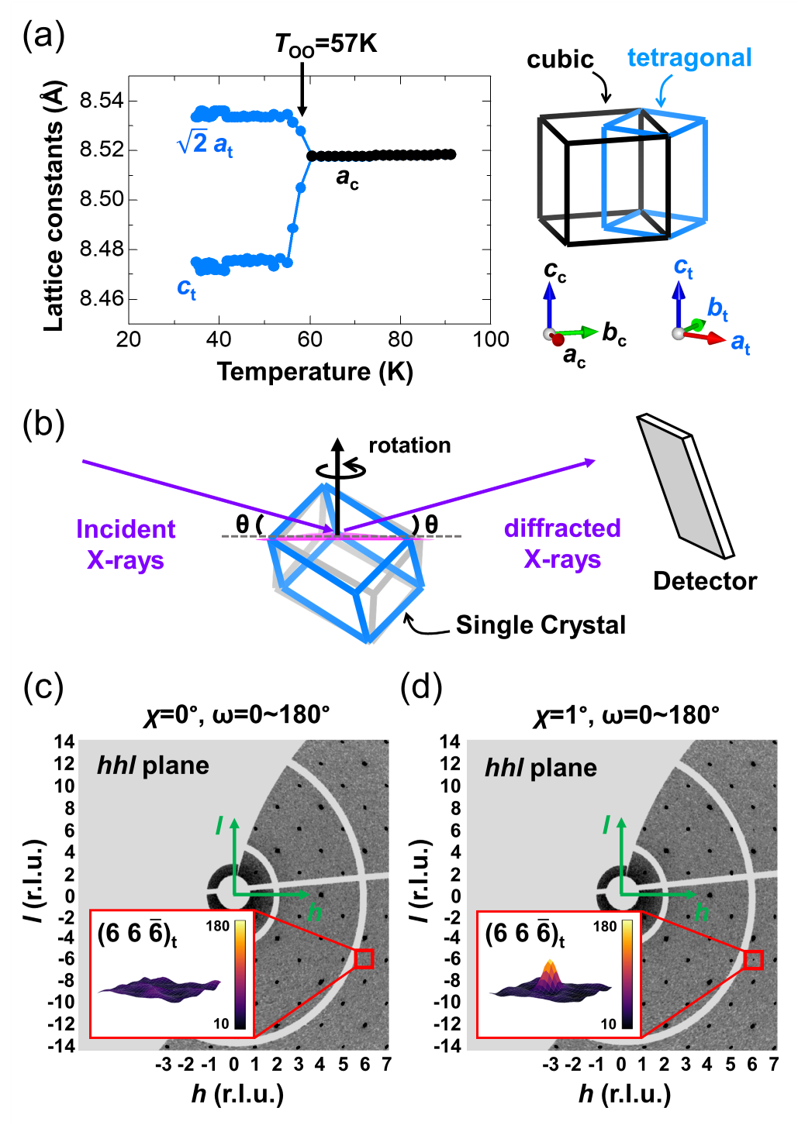}
\caption{(a) Temperature dependence of the lattice constants. A cubic-to-tetragonal phase transition occurs at $T_{\mathrm{OO}} = 57$ K. The relationship between the cubic and tetragonal unit cells is shown in the right panel. (b) Schematic illustration of the experimental concept for distinguishing genuine symmetry-allowed reflections from multiple-scattering artifacts by varying the azimuthal angle of the crystal while keeping the Bragg condition fixed. (c),(d) Reciprocal-space maps in the $hhl$ plane at 40 K in the tetragonal phase, measured at azimuthal angle $\chi = 0^\circ$ and $1^\circ$, respectively. These two-dimensional maps were reconstructed from data collected over an $\omega$ range of $0^\circ$ to $180^\circ$. While the extinction rule associated with $I$ centering ($h+k+l=2n+1$) remains unchanged, reflections corresponding to the $d$-glide extinction rule in the $hhl$ plane ($2h + l \neq 4n$, red squares) appear or disappear upon a slight change in the $\chi$ angle, indicating that they originate from multiple-scattering artifacts. Each inset shows the three-dimensional profile of the respective reflection.}
\label{fig:fig3}
\end{figure}

 Figure~\ref{fig:fig3}(a) shows the temperature dependence of the lattice constant. A structural phase transition from the cubic phase to the tetragonal phase is observed around $T_{\mathrm{OO}}=57$ K. In the previous XRD experiments, the space group of the low-temperature phase has been assigned as $I4_1/a$ based on the violation of the $d$-glide extinction rule~\cite{Suzuki2007,Nii2012}. However, in XRD, multiple scattering can produce intensity at forbidden reflections, necessitating careful discrimination when assessing the true symmetry. Multiple scattering is an accidental process that depends sensitively on the scattering geometry. Therefore, by slightly rotating the crystal around the scattering vector while maintaining the Bragg condition, as shown in Fig.~\ref{fig:fig3}(b), one can distinguish genuine symmetry-allowed reflections from multiple-scattering contributions: the former remain robust against the azimuthal rotation, whereas the latter show a strong dependence on the azimuthal-angle. Figures~\ref{fig:fig3}(c) and \ref{fig:fig3}(d) show reconstructed reciprocal-space maps at 40 K in the $hhl$ plane obtained under varied scattering conditions. For detailed XRD geometry, refer to Fig.S1~\cite{sup}. The intensities of reflections satisfying the $I$-centering condition $h+k+l=2n$ remain unchanged, whereas those forbidden by the $d$-glide condition ($hhl$: $2h+l \neq 4n$, red squares) show large intensity variations under changing scattering conditions [inset of Figs.~\ref{fig:fig3}(c) and (d)]. We therefore conclude that the reflections forbidden by the $d$-glide plane arise from multiple scattering, leading us to assign the proper space group as $I4_1/amd$. Structural refinement performed under this symmetry yields reasonable crystallographic parameters, including atomic displacement parameters (Tables S4-6~\cite{sup}).

\begin{figure}[t]
\centering
\includegraphics[width=\columnwidth]{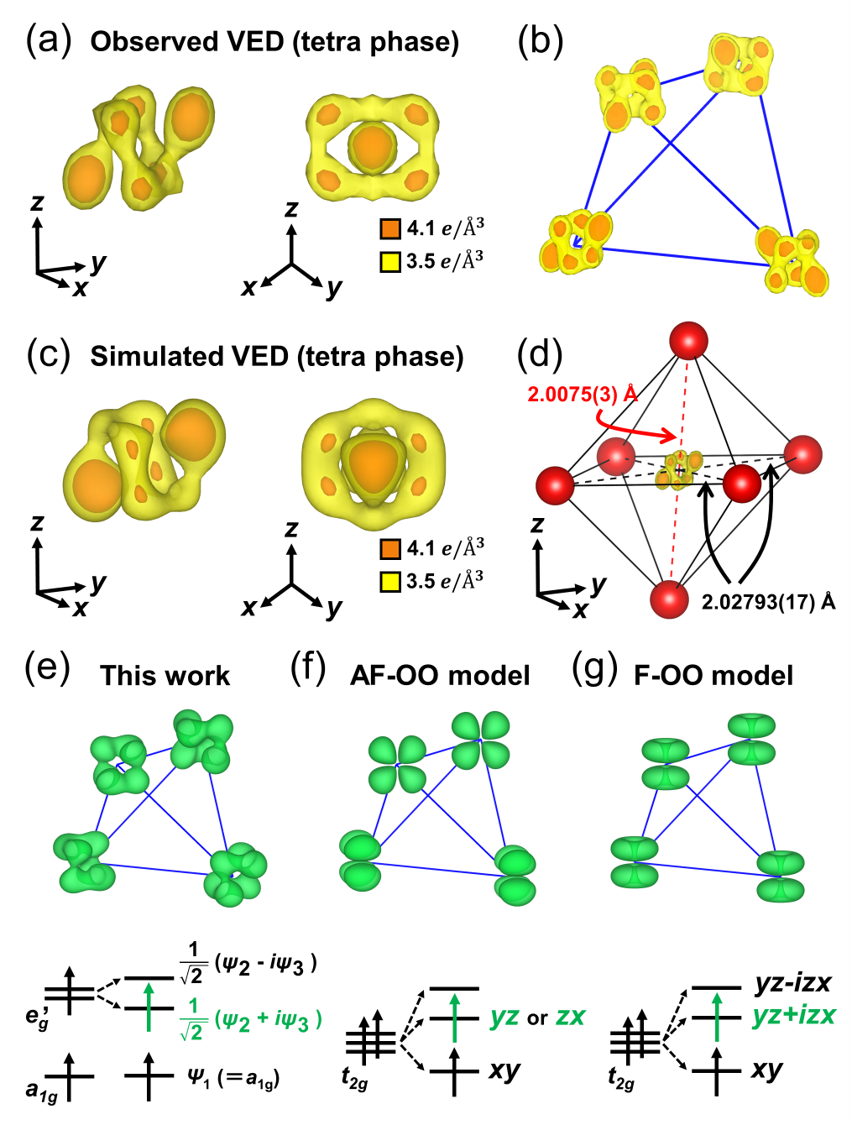}
\caption{(a) VED distribution around the V site at 40 K. Yellow and orange isodensity surfaces represent low and high electron-density levels, respectively. (b) Observed real-space arrangement of the VED on a V$_4$ tetrahedron at 40 K. (c) Simulated VED distributions using quantum parameters obtained by fitting the experimental result. (d) Tetragonally compressed VO$_6$ environment showing the VED around the V site in the tetragonal phase. (e) Real-space arrangement of the electron density corresponding to an electron occupying the $(\psi_2 + i\psi_3)/\sqrt{2}$ orbital, determined from the VED observation. (f),(g) Schematic representations of previously proposed AF-OO and F-OO models. (e), (f), and (g), show the simulated VEDs contributed by the singly occupied orbital indicated by the green arrows in the schematic crystal-field splitting diagrams of the $t_{2g}$ orbitals.}
\label{fig:fig4}
\end{figure}

While the space-group symmetry constrains the possible orbital configurations, it does not uniquely determine the occupied orbital wave function. To resolve this issue, we visualize the VED distribution, which provides direct real-space information on the occupied orbitals. Figures~\ref{fig:fig4}(a) and \ref{fig:fig4}(b) show the VEDs observed around the V site in the low-temperature tetragonal phase and their real-space arrangement on a V$_4$ tetrahedron, respectively. Compared with the VED in the cubic phase [Fig.~\ref{fig:fig2}(a)], the low-temperature distribution in Fig.~\ref{fig:fig4}(a) exhibits two notable changes: the high-density component extending along the $z$ axis is suppressed, and the apparent symmetry of the VED viewed along the $\langle 111\rangle$ axis changes from a nearly sixfold to fourfold-like. The V site symmetry is $..2/m$, which includes a twofold rotation axis along $[1\bar{1}0]$ and a mirror plane perpendicular to this axis. Considering this symmetry, the wave functions of the V $t_{2g}$ orbitals in the tetragonal phase can be written as in Eq.~(1) with $\gamma$ as a variable.

\begin{equation}
\begin{aligned}
\psi_1 &= \sqrt{\frac{1-\gamma^2}{2}}\,\ket{yz}
       + \sqrt{\frac{1-\gamma^2}{2}}\,\ket{zx}
       + \gamma\ket{xy},\\
\psi_2 &= \frac{\gamma}{\sqrt{2}}\ket{yz}
       + \frac{\gamma}{\sqrt{2}}\ket{zx}
       - \sqrt{1-\gamma^2}\,\ket{xy},\\
\psi_3 &= \frac{1}{\sqrt{2}}\ket{yz}
       - \frac{1}{\sqrt{2}}\ket{zx}.
\end{aligned}
\label{eq:eq1}
\end{equation}

Here, $\psi_1$ is the low-temperature counterpart of the $a_{1g}$ orbital identified in the cubic phase, whereas $\psi_2$ and $\psi_3$ correspond to the two components of the residual $e'_g$ doublet. Assuming that the trigonal distortion is large enough for one electron to occupy $\psi_1$, the calculated 3$d^2$ VED is expressed as

\begin{equation}
\rho_{t_{2g}}(\mathbf{r}) = |\psi_1|^2 + \eta |\psi_2|^2 + (1-\eta)|\psi_3|^2,
\end{equation}

To elucidate the V orbital states, the quantum parameters $\gamma$ and $\eta$ are optimized by fitting the calculated VED distributions to reproduce the observed anisotropy. Here, $\gamma$ determines the weight of the $d_{xy}$ component relative to the $d_{yz}$ and $d_{zx}$ components in $\psi_1$ and $\psi_2$, and $\eta$ represents the occupation weight of $\psi_2$ within the residual $e'_g$ doublet. The evaluation function $s$ for the fitting is defined as

\begin{equation}
s = \frac{\sum_{\mathbf{r}} \left|\rho_{\mathrm{obs}}(\mathbf{r}) - \kappa^3 \rho_{t_{2g}}(\kappa \mathbf{r})\right|}{\sum_{\mathbf{r}} |\rho_{\mathrm{obs}}(\mathbf{r})|}.
\end{equation}

Here, $\rho_{\mathrm{obs}}$ is the observed VED around the V site. $\kappa$ is a radial scaling parameter~\cite{Koyama, Kitou2023}. The fitting utilized a VED distribution with a range of $0.25 < r < 0.6$ \AA\ around the V site, where the radial component of the observed VED agrees well with the calculated one (see Supplemental Material~\cite{sup}). The quantum parameters were optimized to $\kappa = 1.21$, $\gamma = 0.74$, and $\eta = 0.49$, for which the evaluation function is minimized to $s = 0.25$ (see Supplemental Material~\cite{sup}). Figure~\ref{fig:fig4}(c) shows the calculated VED distribution using the fitted quantum parameters, which is in good agreement with the experimental result [Fig.~\ref{fig:fig4}(a)]. The fitted value $\gamma = 0.74$ shows that the local orbital state realized in MnV$_2$O$_4$ differs from the orbital configurations assumed in the AF-OO and conventional F-OO models shown in Figs.~\ref{fig:fig4}(f) and (g), both of which correspond to the limiting case $\gamma = 1$. In particular, $\gamma = 0.74$ indicates that the wave function $\psi_1$ is modified from the cubic case ($\gamma = 1/\sqrt{3}$ $\sim 0.58$) and further redistributed toward a larger $d_{xy}$ weight relative to $d_{yz}$ and $d_{zx}$. This imbalance is consistent with the tetragonally compressed VO$_6$ in the low-temperature phase [Fig.~\ref{fig:fig4}(d)]. Within this local orbital state, the fitting yields $\eta = 0.50$, indicating that $\psi_2$ and $\psi_3$ contribute with equal weight to the occupied state. Within a real-orbital description, this result would imply that the $\psi_2$ and $\psi_3$ levels remain in a pseudo-degenerate state. By contrast, once complex orbitals are allowed, the ordered state is naturally interpreted as the complex superposition $(\psi_2\pm i\psi_3)/\sqrt{2}$, selected by relativistic spin-orbit interaction. This interpretation is supported by spin-polarized density-functional-theory calculations with spin--orbit coupling (see Supplemental Material~\cite{sup}). Under the $I4_1/amd$ symmetry, electrons occupying the $(\psi_2 \pm  i\psi_3)/\sqrt{2}$ orbital are arranged on the V$_4$ tetrahedra according to the four differently oriented local axes of the pyrochlore structure [Fig.~4(e)]. Unlike in the AF-OO model, the resulting orbital state retains finite contributions from all three $t_{2g}$ orbitals at each V site, consistent with the substantial interchain exchange inferred from spin-wave analysis~\cite{Chung2008}. The enhanced $d_{xy}$ weight of $\psi_1$ is expected to strengthen antiferromagnetic exchange along V--V bonds lying within the corresponding local $xy$ planes, while weakening the exchange along bonds connecting neighboring planes. This exchange anisotropy is consistent with the two-in-two-out spin configuration of the V sublattice. Through relativistic spin--orbit coupling, the orbital reconfiguration is expected to further stabilize this arrangement. This orbital state is therefore naturally connected to both the noncoplanar 2-in-2-out-type magnetic structure~\cite{Garlea2008,Nakai} and the compressed VO$_6$ geometry. Under the antiferromagnetic Mn--V exchange interaction, the Mn and V moments tend to align antiparallel, while the local anisotropy at the V sites causes the V moments to cant away from a collinear configuration. The magnetic ground state is thus expected to be a noncoplanar ferrimagnetic state.

In summary, we have successfully determined the $t_{2g}$ wave function of V ions in the ground state of MnV$_2$O$_4$ through direct observation of the VED distributions. The resulting orbital state consists of a complex wave function expressed in site-dependent local frames imposed by trigonal distortion. These findings resolve the long-standing ambiguity of the OO state in MnV$_2$O$_4$. Our results thus establish a real-space route to identifying OO states in frustrated materials, opening a way to understand emergent ground states driven by intertwined spin, orbital, and lattice degrees of freedom.

\enlargethispage{\baselineskip}

$Acknowledgments$—We thank H.~Nakai, and K.~Shiratori for fruitful discussions, and A.~Nakano, R.~Misawa, and S.~Aoyagi for supporting the XRD experiment. This work was supported by JST SPRING (Grant No.~JPMJSP2108), JSPS KAKENHI (Grant No.~23H04869, 24H01644, 25H01246, 25H01252, 26H01288, 26H00626, and 26K00649), RIKEN TRIP initiative (Advanced General Intelligence for Science Program, Many-body Electron Systems), JST-ASPIRE (Grant No. JPMJAP2512), and JST FOREST (Grant No.~JPMJFR2362). The synchrotron radiation experiments were performed at SPring-8 with the approval of the Japan Synchrotron Radiation Research Institute (JASRI) (Proposal No.~2023A1520, 2023B2006, 2024B0304, 2025A1714, and 2025A1928).

\vskip\baselineskip

$Data\ availability$—The data that support the findings are available from the authors upon reasonable request.

\end{document}


\setcounter{figure}{0}
\setcounter{table}{0}
\renewcommand{\thefigure}{S\arabic{figure}}
\renewcommand{\thetable}{S\arabic{table}}
{
\centering
{\Large Supplemental information of\par}
\vspace{0.3em}
{\Large \bfseries Observation of complex orbital order in MnV$_2$O$_4$\par}
\vspace{1em}

Chihaya Koyama,$^{1,*}$ Shunsuke Kitou,$^{1,\dagger}$ Taishun Manjo,$^{2}$ Yuiga Nakamura,$^{2}$ Takeshi Hara,$^{3}$\par
Naoyuki Katayama,$^{4}$ Takuro Katsufuji,$^{5,6}$ Ryotaro Arita,$^{7,8}$ Yusuke Nomura,$^{9,10}$\par
Hiroshi Sawa,$^{11}$ and Taka-hisa Arima$^{1,8}$\par

\vspace{0.8em}

{\small\itshape
$^{1}$ Department of Advanced Materials Science, The University of Tokyo, Kashiwa 277-8561, Japan.\par
$^{2}$ Japan Synchrotron Radiation Research Institute (JASRI), SPring-8, Hyogo 679-5198, Japan.\par
$^{3}$ Department of Physics, Tohoku University, Sendai 980-8578, Japan.\par
$^{4}$ Department of Physics, Okayama University, Okayama 700-8530, Japan.\par
$^{5}$ Department of Physics, Waseda University, Tokyo 169-8555, Japan.\par
$^{6}$ Kagami Memorial Research Institute for Materials Science and Technology, Waseda University, Tokyo 169-0051, Japan.\par
$^{7}$ Department of Physics, The University of Tokyo, Tokyo 153-8904, Japan.\par
$^{8}$ RIKEN Center for Emergent Matter Science (CEMS), Wako 351-0198, Japan.\par
$^{9}$ Institute for Materials Research (IMR), Tohoku University, Sendai 980-8577, Japan.\par
$^{10}$ International Quantum Materials Center (Q²), Advanced Institute for Materials Research (WPI-AIMR), Tohoku University, Sendai 980-8577, Japan.\par
$^{11}$ Nagoya Industrial Science Research Institute, Nagoya 460-0008, Japan.\par
}
\par
}

\vspace{1em}

\noindent $^{*}$koyama-chihaya@g.ecc.u-tokyo.ac.jp\\
\noindent $^{\dagger}$kitou@edu.k.u-tokyo.ac.jp

\vspace{1em}

\textbf{This PDF file includes:}

Supplementary text

Figures S1 to S7

Tables S1 to S7

SI References

\vspace{1em}

\textbf{CONTENTS}

1. Single-crystal structural analysis using synchrotron X-ray
diffraction......................2

2. Experimental verification of multiple scattering in the tetragonal phase..................5

3. Valence electron density analysis..............................................................................7

4. Spin-polarized DFT+\emph{U}+SOC
calculations\ldots..........................................................9

\vspace{1em}

\clearpage
\textbf{1.} \textbf{Single-crystal structural analysis using synchrotron
X-ray diffraction.}

\hspace*{1em} MnV$_2$O$_4$ undergoes structural phase
transition---from the cubic phase to the 
tetragonal phase upon cooling, as described in the main
manuscript. The structural analysis results for
MnV$_2$O$_4$ at 160 K (cubic) and 40 K (tetragonal) are summarized in
\hyperref[Table_S1]{Tables S1-6}.

\vspace{1em}

\phantomsection\label{Table_S1}{}Table S1. Summary of crystallographic
data of MnV$_2$O$_4$ (cubic phase).

\begin{table}[ht]
\centering
\label{tab:S1}
\setlength{\tabcolsep}{6pt}
\begin{tabular}{c c}
\hline
Temperature (K) & 160 \\
Wavelength $\lambda$ (\AA) & 0.30945 \\
Crystal dimension ($\mu$m$^3$) & $40 \times 40 \times 40$ \\
Space group & $Fd\bar{3}m$ \\
$a$ (\AA) & 8.5217(10) \\
$V$ (\AA$^3$) & 618.840(2) \\
$Z$ & 8 \\
$(\sin\theta/\lambda)_{\max}$ (\AA$^{-1}$) & 1.79 \\
$N_{\mathrm{Total,obs}}$ & 30923 \\
$N_{\mathrm{Unique,obs}}$ ($I>3\sigma$ / all) & 713 / 751 \\
Average redundancy & 38.6 \\
Completeness & 0.996 \\
$R_{\mathrm{int}}$ & 0.042 \\
\hline
\multicolumn{2}{l}{Only high-angle reflections $[0.8 \le \sin\theta/\lambda \le 1.79$ \AA$^{-1}]$} \\
\hline
$R_1$ ($I>3\sigma$ / all) & 0.0088 / 0.0094 \\
$wR$ ($I>3\sigma$ / all) & 0.0172 / 0.0182 \\
GOF ($I>3\sigma$ / all) & 0.99 / 1.01 \\
\hline
\multicolumn{2}{l}{All reflections $[0 \le \sin\theta/\lambda \le 1.79$ \AA$^{-1}]$} \\
\hline
$R_1$ ($I>3\sigma$ / all) & 0.0115 / 0.0120 \\
$wR$ ($I>3\sigma$ / all) & 0.0225 / 0.0231 \\
GOF ($I>3\sigma$ / all) & 1.30 / 1.30 \\
\hline
\end{tabular}
\end{table}

\vspace{1em}

Table S2. Structural parameters of
MnV$_2$O$_4$ at 160 K using origin choice 2.

\begin{longtable}[]{@{}
  >{\raggedright\arraybackslash}p{(\columnwidth - 10\tabcolsep) * \real{0.0912}}
  >{\raggedright\arraybackslash}p{(\columnwidth - 10\tabcolsep) * \real{0.1245}}
  >{\raggedright\arraybackslash}p{(\columnwidth - 10\tabcolsep) * \real{0.2015}}
  >{\raggedright\arraybackslash}p{(\columnwidth - 10\tabcolsep) * \real{0.2015}}
  >{\raggedright\arraybackslash}p{(\columnwidth - 10\tabcolsep) * \real{0.2015}}
  >{\raggedright\arraybackslash}p{(\columnwidth - 10\tabcolsep) * \real{0.1799}}@{}}
\toprule\noalign{}
\begin{minipage}[b]{\linewidth}\raggedright
Atom
\end{minipage} & \begin{minipage}[b]{\linewidth}\raggedright
Wyckoff position
\end{minipage} & \begin{minipage}[b]{\linewidth}\raggedright
\emph{x}
\end{minipage} & \begin{minipage}[b]{\linewidth}\raggedright
\emph{y}
\end{minipage} & \begin{minipage}[b]{\linewidth}\raggedright
\emph{z}
\end{minipage} & \begin{minipage}[b]{\linewidth}\raggedright
\emph{U\textsubscript{eq}} ($\AA$\textsuperscript{2})
\end{minipage} \\
\midrule\noalign{}
\endhead
\bottomrule\noalign{}
\endlastfoot
Mn & 8\emph{a} & \(1/8\) & \(1/8\) & \(1/8\) & \(0.003379(8)\) \\
V & 16\emph{d} & \(1/2\) & \(1/2\) & \(1/2\) & \(0.002724(8)\) \\
O & 32\emph{e} & \(0.737133(11)\) & \(0.737133(11)\) & \(0.737133(11)\)
& \(0.004290(9)\) \\
\end{longtable}

Table S3. Anisotropic atomic displacement parameters of
MnV$_2$O$_4$ at 160 K.

\begin{longtable}[]{@{}
  >{\raggedright\arraybackslash}p{(\columnwidth - 12\tabcolsep) * \real{0.0699}}
  >{\raggedright\arraybackslash}p{(\columnwidth - 12\tabcolsep) * \real{0.1435}}
  >{\raggedright\arraybackslash}p{(\columnwidth - 12\tabcolsep) * \real{0.1435}}
  >{\raggedright\arraybackslash}p{(\columnwidth - 12\tabcolsep) * \real{0.1435}}
  >{\raggedright\arraybackslash}p{(\columnwidth - 12\tabcolsep) * \real{0.1665}}
  >{\raggedright\arraybackslash}p{(\columnwidth - 12\tabcolsep) * \real{0.1665}}
  >{\raggedright\arraybackslash}p{(\columnwidth - 12\tabcolsep) * \real{0.1665}}@{}}
\toprule\noalign{}
\begin{minipage}[b]{\linewidth}\raggedright
Atom
\end{minipage} & \begin{minipage}[b]{\linewidth}\raggedright
\emph{U}\textsubscript{11} ($\AA$\textsuperscript{2})
\end{minipage} & \begin{minipage}[b]{\linewidth}\raggedright
\emph{U}\textsubscript{22} ($\AA$\textsuperscript{2})
\end{minipage} & \begin{minipage}[b]{\linewidth}\raggedright
\emph{U}\textsubscript{33} ($\AA$\textsuperscript{2})
\end{minipage} & \begin{minipage}[b]{\linewidth}\raggedright
\emph{U}\textsubscript{12} ($\AA$\textsuperscript{2})
\end{minipage} & \begin{minipage}[b]{\linewidth}\raggedright
\emph{U}\textsubscript{13} ($\AA$\textsuperscript{2})
\end{minipage} & \begin{minipage}[b]{\linewidth}\raggedright
\emph{U}\textsubscript{23} ($\AA$\textsuperscript{2})
\end{minipage} \\
\midrule\noalign{}
\endhead
\bottomrule\noalign{}
\endlastfoot
Mn & \(0.003379(13)\) & \(= U_{11}\) & \(= U_{11}\) & \(0\) & \(0\) &
\(0\) \\
V & \(0.002724(13)\) & \(= U_{11}\) & \(= U_{11}\) & \(-0.000144(3)\) &
\(= U_{12}\) & \(= U_{12}\) \\
O & \(0.004290(16)\) & \(= U_{11}\) & \(= U_{11}\) & \(-0.000398(14)\) &
\(= U_{12}\) & \(= U_{12}\) \\
\end{longtable}

\clearpage

Table S4. Summary of crystallographic data of
MnV$_2$O$_4$ (tetragonal phase).

\begin{table}[ht]
\centering
\label{tab:S4}
\setlength{\tabcolsep}{6pt}
\begin{tabular}{c c}
\hline
Temperature (K) & 40 \\
Wavelength $\lambda$ (\AA) & 0.30945 \\
Crystal dimension ($\mu$m$^3$) & $40 \times 40 \times 40$ \\
Space group & $I4_1/amd$ \\
$a$ (\AA) & 6.0313(2) \\
$c$ (\AA) & 8.4339(3) \\
$V$ (\AA$^3$) & 306.796(18) \\
$Z$ & 4 \\
$(\sin\theta/\lambda)_{\max}$ (\AA$^{-1}$) & 1.79 \\
$N_{\mathrm{Total,obs}}$ & 35850 \\
$N_{\mathrm{Unique,obs}}$ ($I>3\sigma$ / all) & 1618 / 1922 \\
Average redundancy & 18.3 \\
Completeness & 0.980 \\
$R_{\mathrm{int}}$ & 0.037 \\
\hline
\multicolumn{2}{l}{Only high-angle reflections $[0.8 \le \sin\theta/\lambda \le 1.79$ \AA$^{-1}]$} \\
\hline
$R_1$ ($I>3\sigma$ / all) & 0.0145 / 0.0192 \\
$wR$ ($I>3\sigma$ / all) & 0.0235 / 0.0246 \\
GOF ($I>3\sigma$ / all) & 1.17 / 1.11 \\
\hline
\multicolumn{2}{l}{All reflections $[0 \le \sin\theta/\lambda \le 1.79$ \AA$^{-1}]$} \\
\hline
$R_1$ ($I>3\sigma$ / all) & 0.0153 / 0.0190 \\
$wR$ ($I>3\sigma$ / all) & 0.0251 / 0.0261 \\
GOF ($I>3\sigma$ / all) & 1.25 / 1.19 \\
\hline
\end{tabular}
\end{table}

\vspace{1em}

Table S5. Structural parameters of
MnV$_2$O$_4$ at 40K using origin choice 2.

\begin{longtable}[]{@{}
  >{\raggedright\arraybackslash}p{(\columnwidth - 10\tabcolsep) * \real{0.0912}}
  >{\raggedright\arraybackslash}p{(\columnwidth - 10\tabcolsep) * \real{0.1245}}
  >{\raggedright\arraybackslash}p{(\columnwidth - 10\tabcolsep) * \real{0.2015}}
  >{\raggedright\arraybackslash}p{(\columnwidth - 10\tabcolsep) * \real{0.2015}}
  >{\raggedright\arraybackslash}p{(\columnwidth - 10\tabcolsep) * \real{0.2015}}
  >{\raggedright\arraybackslash}p{(\columnwidth - 10\tabcolsep) * \real{0.1799}}@{}}
\toprule\noalign{}
\begin{minipage}[b]{\linewidth}\raggedright
Atom
\end{minipage} & \begin{minipage}[b]{\linewidth}\raggedright
Wyckoff position
\end{minipage} & \begin{minipage}[b]{\linewidth}\raggedright
\emph{x}
\end{minipage} & \begin{minipage}[b]{\linewidth}\raggedright
\emph{y}
\end{minipage} & \begin{minipage}[b]{\linewidth}\raggedright
\emph{z}
\end{minipage} & \begin{minipage}[b]{\linewidth}\raggedright
\emph{U\textsubscript{eq}} ($\AA$\textsuperscript{2})
\end{minipage} \\
\midrule\noalign{}
\endhead
\bottomrule\noalign{}
\endlastfoot
Mn & 4\emph{a} & \(0\) & \(3/4\) & \(1/8\) & \(0.001962(8)\) \\
V & 8\emph{d} & \(0\) & \(0\) & \(1/2\) & \(0.001643(9)\) \\
O & 16\emph{h} & \(0\) & \(0.47415(4)\) & \(0.26269(3)\) &
\(0.00336(2)\) \\
\end{longtable}

Table S6. Anisotropic atomic displacement parameters of
MnV$_2$O$_4$ at 40 K.

\begin{longtable}[]{@{}
  >{\raggedright\arraybackslash}p{(\columnwidth - 12\tabcolsep) * \real{0.0833}}
  >{\raggedright\arraybackslash}p{(\columnwidth - 12\tabcolsep) * \real{0.1774}}
  >{\raggedright\arraybackslash}p{(\columnwidth - 12\tabcolsep) * \real{0.1774}}
  >{\raggedright\arraybackslash}p{(\columnwidth - 12\tabcolsep) * \real{0.1774}}
  >{\raggedright\arraybackslash}p{(\columnwidth - 12\tabcolsep) * \real{0.1105}}
  >{\raggedright\arraybackslash}p{(\columnwidth - 12\tabcolsep) * \real{0.1105}}
  >{\raggedright\arraybackslash}p{(\columnwidth - 12\tabcolsep) * \real{0.1635}}@{}}
\toprule\noalign{}
\begin{minipage}[b]{\linewidth}\raggedright
Atom
\end{minipage} & \begin{minipage}[b]{\linewidth}\raggedright
\emph{U}\textsubscript{11} ($\AA$\textsuperscript{2})
\end{minipage} & \begin{minipage}[b]{\linewidth}\raggedright
\emph{U}\textsubscript{22} ($\AA$\textsuperscript{2})
\end{minipage} & \begin{minipage}[b]{\linewidth}\raggedright
\emph{U}\textsubscript{33} ($\AA$\textsuperscript{2})
\end{minipage} & \begin{minipage}[b]{\linewidth}\raggedright
\emph{U}\textsubscript{12} ($\AA$\textsuperscript{2})
\end{minipage} & \begin{minipage}[b]{\linewidth}\raggedright
\emph{U}\textsubscript{13} ($\AA$\textsuperscript{2})
\end{minipage} & \begin{minipage}[b]{\linewidth}\raggedright
\emph{U}\textsubscript{23} ($\AA$\textsuperscript{2})
\end{minipage} \\
\midrule\noalign{}
\endhead
\bottomrule\noalign{}
\endlastfoot
Mn & \(0.001963(13)\) & \(= U_{11}\) & \(0.001960(16)\) & \(0\) & \(0\) &
\(0\) \\
V & \(0.001733(15)\) & \(0.001553(15)\) & \(0.001644(15)\) & \(0\) &
\(0\) & \(0.000136(8)\) \\
O & \(0.00439(5)\) & \(0.00279(4)\) & \(0.00291(4)\) & \(0\) & \(0\) &
\(0.00036(3)\) \\
\end{longtable}

\clearpage

\hspace*{1em} In the tetragonal phase, the single crystal used in this study contained two twin domains characterized by the following orientation matrices: 
\begin{equation}
\begin{pmatrix}
h\\
k\\
l
\end{pmatrix}_{\mathrm{twin}\,2}
=
\begin{pmatrix}
1/2 & 1/2 & -1/2 \\
1/2 & 1/2 &  1/2 \\
1   & -1  &  0
\end{pmatrix}
\begin{pmatrix}
h\\
k\\
l
\end{pmatrix}_{\mathrm{twin}\,1}.
\end{equation}
The refined volume fractions for the tetragonal phase are approximately 0.5790(2) and 0.4210(2). By collecting multiple diffraction datasets that comprehensively cover reciprocal space, we identified regions in which all equivalent reflections originated exclusively from the major domain. The combined datasets provided sufficient completeness and redundancy, enabling reliable structural refinement and CDFS analysis using only reflections from the major domain.Only these reflections, which provide the most reliable diffraction intensities, were used for structural refinements and CDFS analyses.

\begin{figure}[htbp]
\centering
\includegraphics[width=\columnwidth]{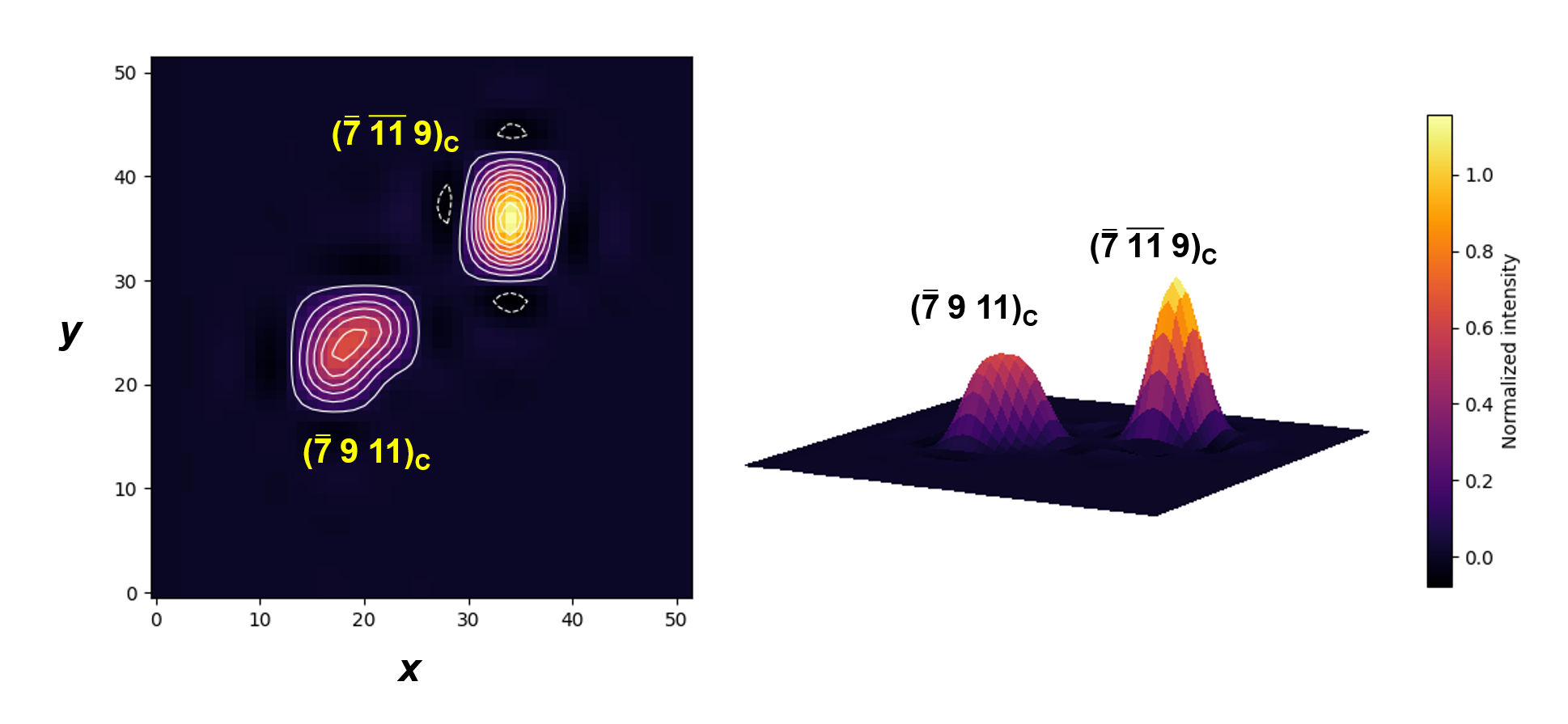}
\caption{Heat map (left) and three-dimensional plot of normalized diffraction intensity (right) for two domain-related reflections in the tetragonal phase of MnV$_2$O$_4$ at 40~K. Both reflections are indexed based on the parent cubic setting. The intensity distribution illustrates the relative volume fractions of the structural domains. The stronger reflection, corresponding to $(\bar{7}\,\overline{11}\,9)_{\rm c}$, originates from the major domain, whereas the weaker reflection, $(\bar{7}\,9\,11)_{\rm c}$, originates from the minor domain. The cubic-to-tetragonal transition can generate up to three tetragonal structural domains, whereas only two components were observed in this measurement; the remaining component expected at $(9\,\overline{11}\,7)_{\rm c}$ was absent.}
\label{fig:twin}
\end{figure}

\clearpage

\textbf{2. Experimental verification of multiple scattering in the tetragonal phase.}

\begin{figure}[htbp]
\centering
\includegraphics[width=\columnwidth]{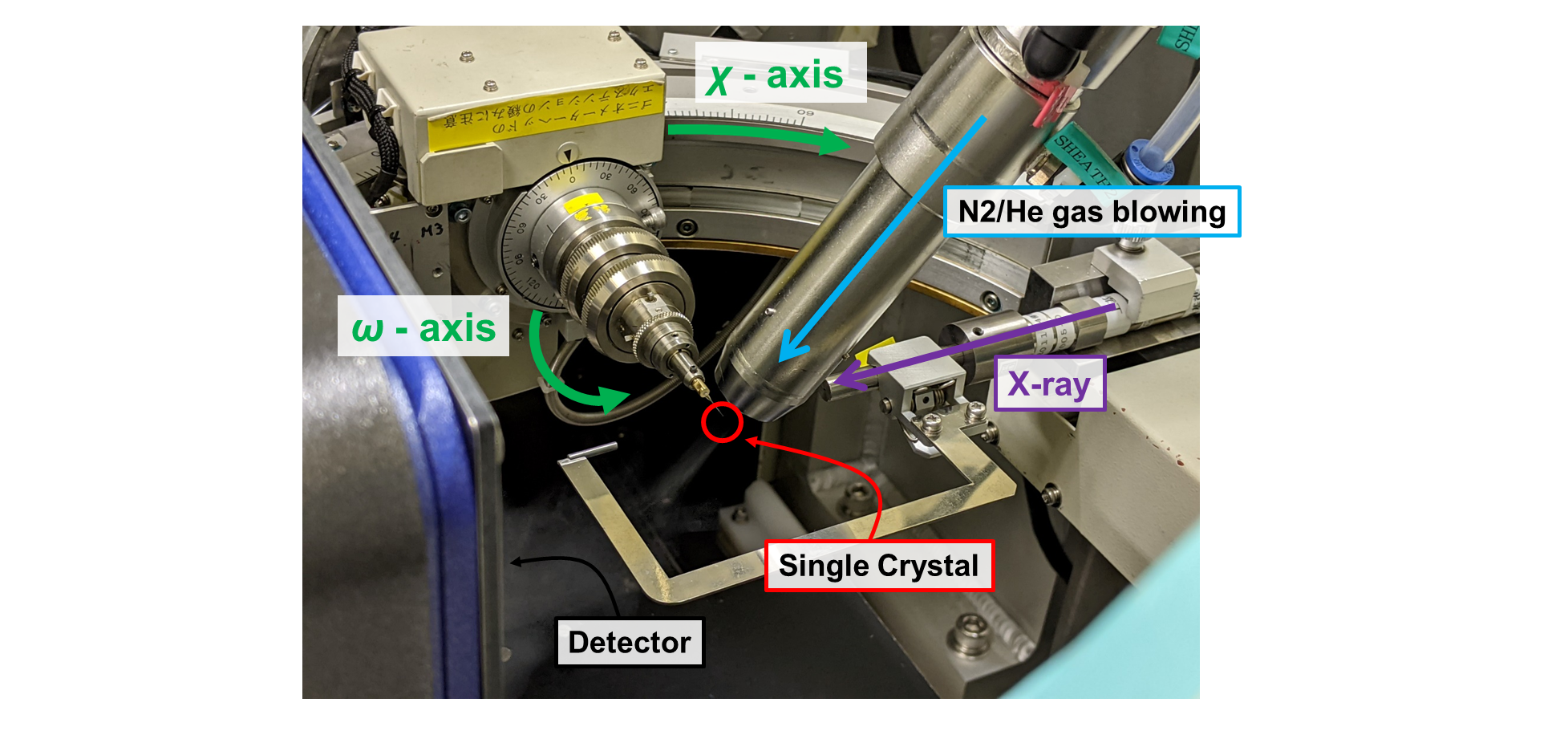}
\caption{Set-up for single-crystal X-ray diffraction experiments. Diffraction data were acquired by rotating $\omega$ from $0^\circ$ to $180^\circ$ at each fixed $\chi$ angle, with $\chi$ varied in $1^\circ$ increments. }
\label{fig:setup}
\end{figure}

\begin{figure}[htbp]
\centering
\includegraphics[width=\columnwidth]{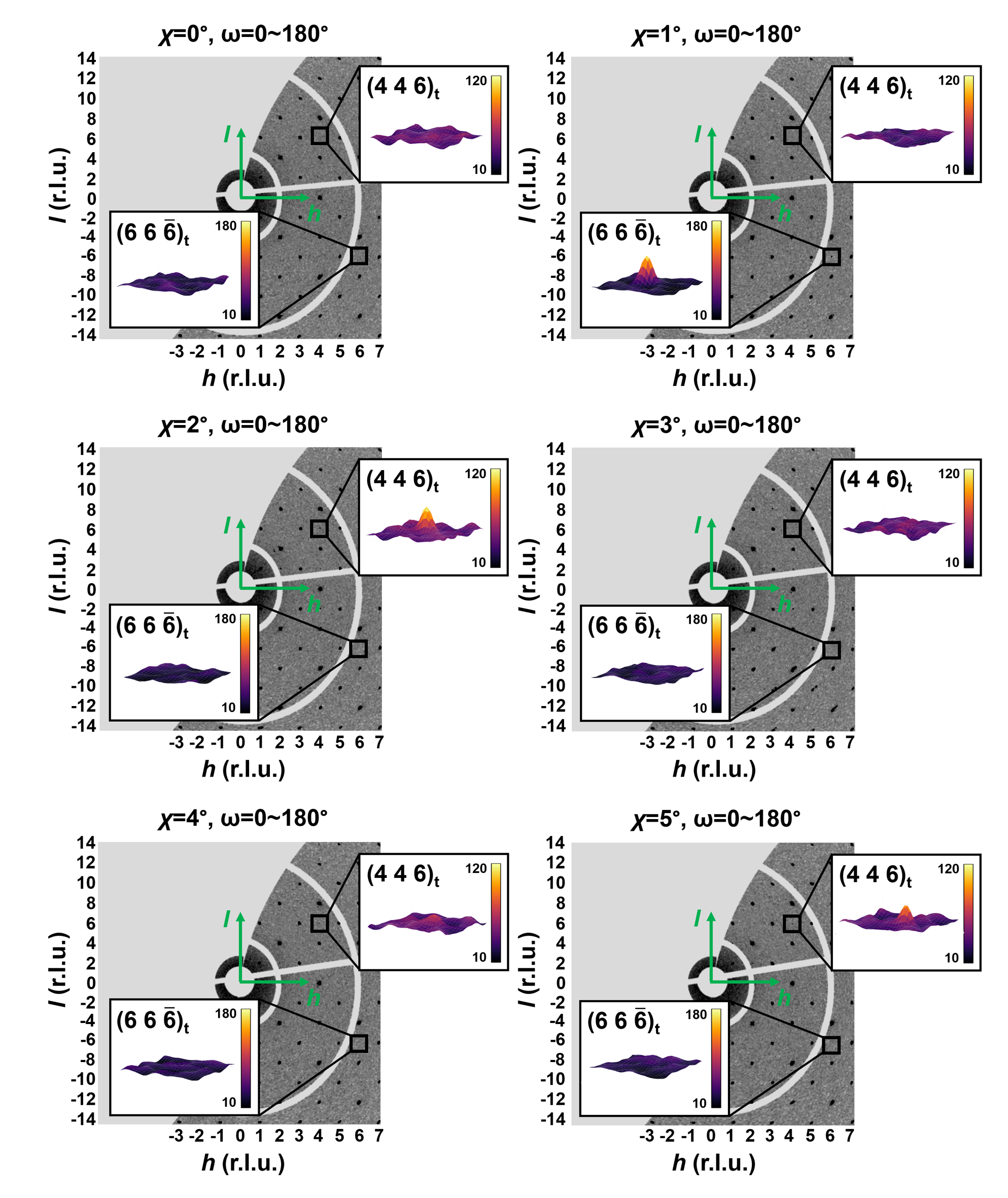}
\caption{Reciprocal-space maps in the $hhl$ plane at 40 K in the tetragonal phase, measured at azimuthal angles $\chi = 0^\circ$, $1^\circ$, $2^\circ$, $3^\circ$, $4^\circ$, and $5^\circ$. These two-dimensional maps were reconstructed from data collected over an $\omega$ range of $0^\circ$ to $180^\circ$. }
\label{fig:hhl}
\end{figure}

\hspace*{1em} Figure~\ref{fig:setup} shows the detailed geometry of the diffractometer. To clarify the long-standing controversy over whether the low-temperature space group is $I4_1/a$ or $I4_1/amd$, we carried out azimuth-dependent diffraction measurements. We focused on reflections violating the $d$-glide extinction condition in the $hhl$ plane, namely $hhl$ with $2h+l \neq 4n$, because these reflections have been the central basis for the assignment of $I4_1/a$ in previous studies. From the scattering-geometry dependence of these forbidden reflections, we distinguished genuine symmetry breaking from multiple-scattering artifacts. Figure~\ref{fig:hhl} shows the reconstructed reciprocal-space map in the $hhl$ planes at 40 K in the tetragonal phase.


\clearpage

\clearpage

\textbf{3.} \textbf{Valence electron density analysis}

\begin{figure}[htbp]
\centering
\includegraphics[width=\columnwidth]{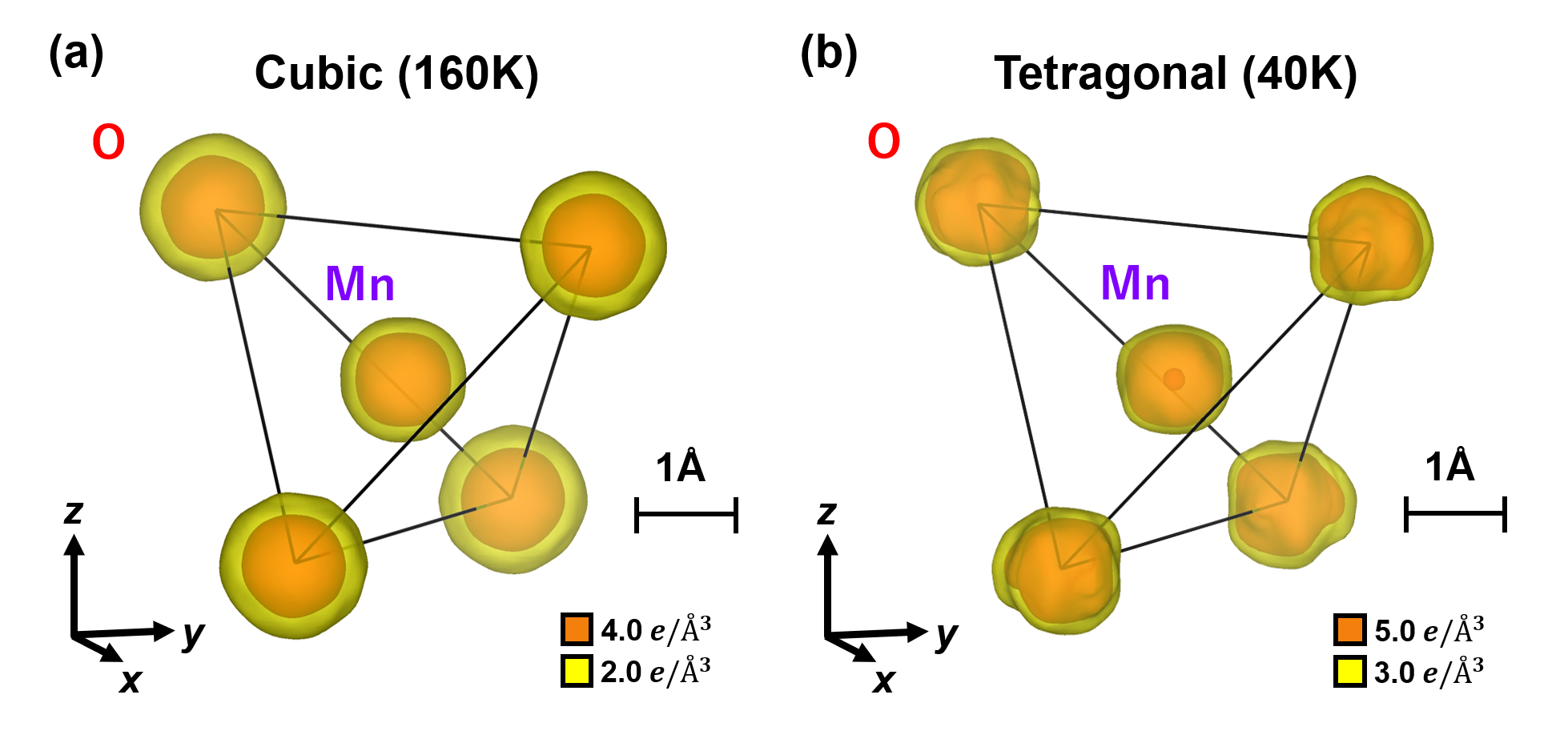}
\caption{(a) Experimentally obtained VED distribution around Mn and O sites in the cubic phase. The iso-density surfaces at 2.0 e/\AA$^3$ and 4.0 e/\AA$^3$ are displayed in yellow and orange, respectively. (b) Experimentally obtained VED distribution around Mn and O sites in the tetragonal phase. The iso-density surfaces at 3.0 e/\AA$^3$ and 5.0 e/\AA$^3$ are displayed in yellow and orange, respectively. }
\label{fig:Mn_VED}
\end{figure}

\hspace*{1em} Figure~\ref{fig:Mn_VED} shows the VED distributions around the Mn and O sites in the cubic and tetragonal phases, as obtained by the CDFS method. The VED distribution around the O site is nearly spherical, consistent with the closed-shell O$^{2-}$ electron configuration ($2s^2 2p^6$). 
Similarly, because Mn$^{2+}$ ($3d^5$) in the high-spin state has no active orbital degree of freedom, the Mn $3d$ orbitals are uniformly occupied. Accordingly, the VED distribution around the Mn site is expected to be isotropic. The structural phase transition from the cubic to tetragonal phase does not induce significant distortion of the MnO$_4$ tetrahedron (Table~\ref{tab:mno4_angles}), and the electronic state at the Mn site remains essentially unchanged, as shown in Fig.~\ref{fig:Mn_VED}(b).

\clearpage

Table S7. O--Mn--O bond angles of the MnO$_4$ tetrahedra in MnV$_2$O$_4$. The values were calculated from the structural parameters of the cubic phase at 160~K and the tetragonal phase at 40~K.

\setcounter{table}{6}
\refstepcounter{table}
\label{tab:mno4_angles}
\begin{center}
\begin{tabular}{ccccc}
\hline
Phase & Temperature (K) & O--Mn--O angle (deg.) & Multiplicity & $\Delta\theta$ (deg.) \\
\hline
cubic & 160 & 109.47 & 6 & -- \\
\multirow[c]{2}{2.0cm}{\centering tetragonal} & \multirow[c]{2}{1.2cm}{\centering 40} & 109.12 & 4 & $-0.35$ \\
 & & 110.17 & 2 & $+0.70$ \\
\hline
\end{tabular}
\end{center}

\vspace{5em}

\begin{figure}[htbp]
\centering
\includegraphics[width=\columnwidth]{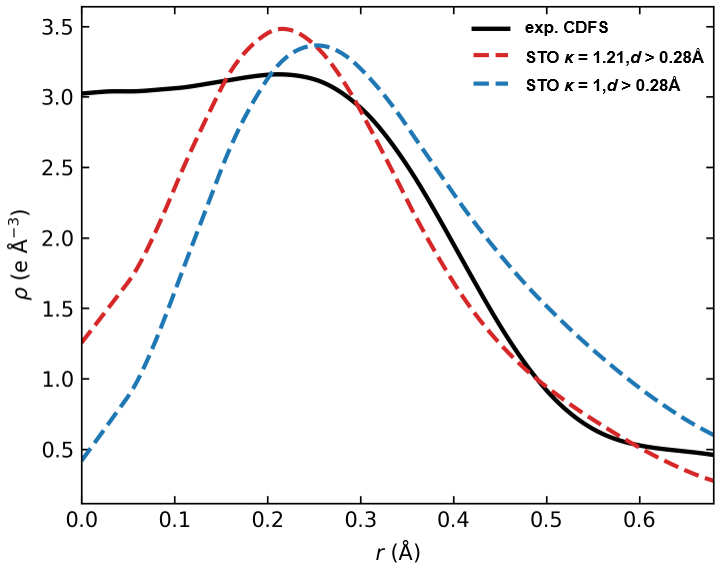}
\caption{Radial profiles of the VED around the V site.
The black solid line represents the experimentally obtained radial component.
The red and blue dashed lines show profiles calculated using Slater-type orbitals (STO)~\cite{Su1997,Macchi2001} for an isolated V atom with radial scaling parameters of $\kappa = 1.21$ and $\kappa = 1.00$, respectively.
For comparison with the experimental CDFS map, the calculated STO profiles were Fourier-filtered using the experimental resolution cutoff of $d>0.28~\mathrm{\AA}$~\cite{Kitou2023}.}
\label{fig:radial_profile}
\end{figure}

\clearpage

\textbf{4. Spin-polarized} \textbf{DFT+\emph{U}+SOC calculations.}

\hspace*{1em} The DFT electronic structure calculations for MnV$_2$O$_4$ were performed using Quantum Espresso~\cite{Giannozzi2017} with the experimental structure of the tetragonal phase obtained in this study. We employed relativistic norm-conserving pseudopotentials with Perdew-Burke-Ernzerhof (PBE)~\cite{Perdew1996} exchange-correlation functional, which were taken from the PseudoDojo~\cite{vanSetten2018}. We performed spin-polarized calculations considering the spin-orbit coupling with $\boldsymbol{k}$-mesh of 7×7×7. The effects of Coulomb interactions of Mn 3$d$ and V 3$d$ orbitals were incorporated within the DFT+$U$ formalism~\cite{Liechtenstein1995} with the Hubbard $U$ parameters of 5 eV. The energy cutoff was set to 100 Ry for the wave functions and 400 Ry for the charge density. Based on the DFT electronic structure, we constructed Wannier orbitals~\cite{Marzari1997, Souza2002} using RESPACK~\cite{Nakamura2021,Charlebois2021} and Wannier90~\cite{Pizzi2020} for Mn 3$d$, V 3$d$, and O 2$p$ manifold. Given that there are two Mn sites, four V sites, and eight O sites within the primitive unit cell, the number of Wannier orbitals was (5×2+5×4+3×8)×2=108, where the factor of two accounted for the summation over spin degrees of freedom.

\begin{figure}[htbp]
\centering
\includegraphics[width=\columnwidth]{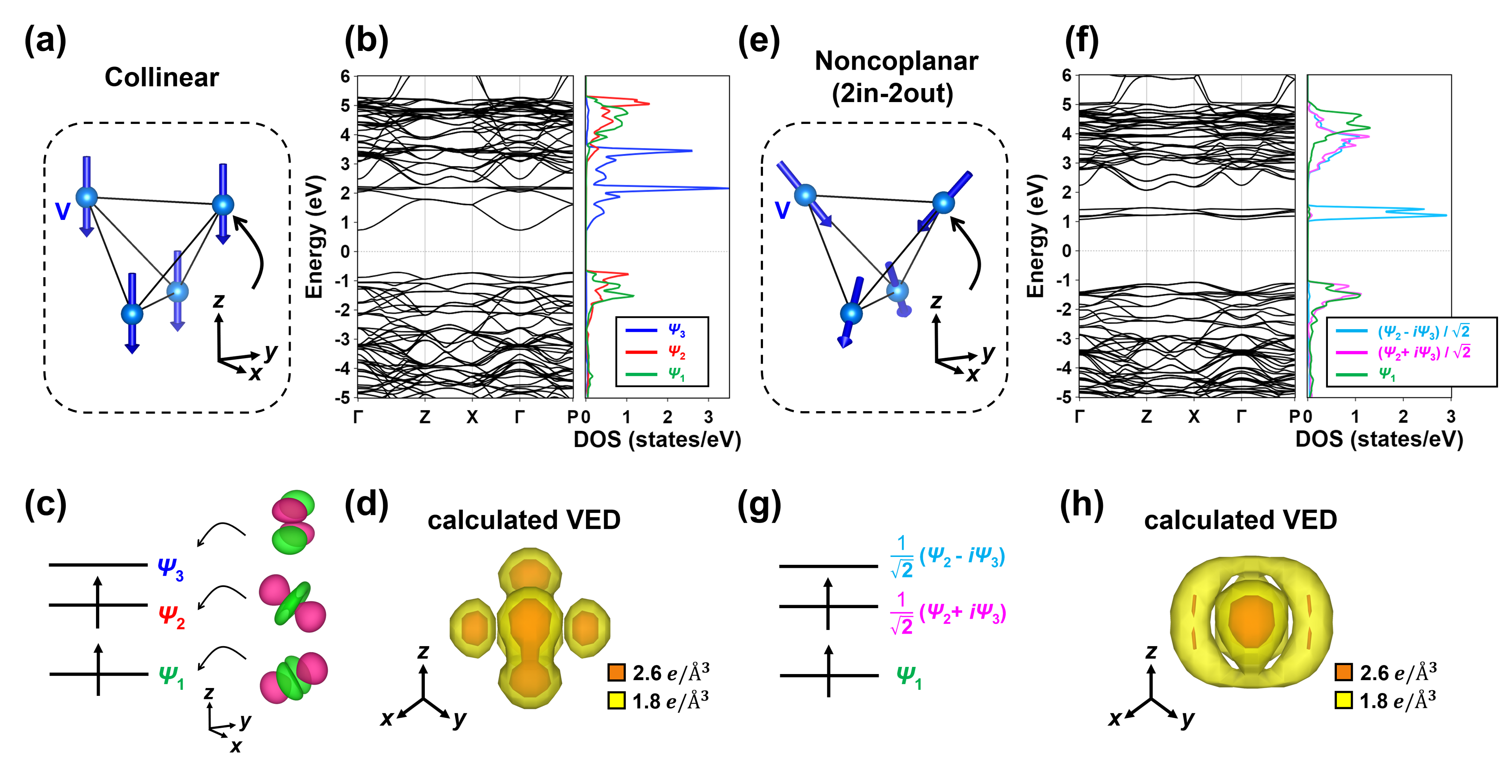}
\caption{Results of spin-polarized DFT+$U$ calculations including spin--orbit coupling for the tetragonal phase. (a) Magnetic structure in the nearly collinear solution, where the V spins are antiferromagnetically aligned with the Mn spins. (b) Band structure along the $\boldsymbol{k}$-path $\Gamma(0,0,0)\text{--} Z(2\pi/a,0,0)\text{--} X(\pi/a,\pi/a,0)\text{--} \Gamma(0,0,0)\text{--} P(\pi/a,\pi/a,\pi/c)$ and projected density of states of the V $t_{2g}$ orbitals for the nearly collinear solution. (c), (d) Schematic diagram of the orbital states corresponding to the nearly collinear solution and the calculated VED distribution at the V site. (e) Magnetic structure in the noncoplanar solution with two-in--two-out-type V magnetic moments. In this solution, the V spin is canted by approximately $54^\circ$ from the $-z$ direction. (f) Band structure and projected density of states of the V $t_{2g}$ orbitals for the noncoplanar solution. (g), (h) Schematic diagram of the orbital states corresponding to the noncoplanar solution and the calculated VED distribution at the V site.}
\label{fig:DFT}
\end{figure}

\hspace*{1em} In understanding orbital ordering in the V $t_{2g}$ subspace in the tetragonal phase, it is important to keep in mind the energy scales of crystal-field splitting and spin--orbit coupling. The crystal-field splitting induced by the $D_{3d}$-type distortion provides the largest energy scale and lifts the degeneracy of the $t_{2g}$ orbitals into the low-energy $a_{1g}$ orbital ($\psi_1$) and the higher-energy $e_g^{\prime}$ orbitals ($\psi_2$ and $\psi_3$). Smaller energy scales are associated with the additional crystal-field splitting caused by $c$-axis compression and with spin--orbit coupling. Since the splitting into the $a_{1g}$ and $e_g^{\prime}$ orbitals is already present in the cubic phase, the key questions are how orbital order is formed from $\psi_2$ and $\psi_3$, and which interaction plays the decisive role in selecting the ordered state.\\
\hspace*{1em} To address this issue, we perform spin-polarized DFT+$U$ calculations including spin--orbit coupling. The three V $t_{2g}$ wavefunctions are defined as $\psi_1 = \sqrt{(1-\gamma^2)/2}\,|yz\rangle + \sqrt{(1-\gamma^2)/2}\,|zx\rangle + \gamma|xy\rangle$,
$\psi_2 = (\gamma/\sqrt{2})|yz\rangle + (\gamma/\sqrt{2})|zx\rangle -\sqrt{1-\gamma^2}\,|xy\rangle$, and
$\psi_3 = (1/\sqrt{2})|yz\rangle - (1/\sqrt{2})|zx\rangle$ [as shown in Fig. ~\ref{fig:DFT}(c)]. The Hubbard $U$ lowers the energies of occupied states and raises those of unoccupied states, thereby promoting gap formation. Interestingly, we find two solutions depending on the mechanism of orbital order.\\
\hspace*{1em} One solution is a nearly collinear ferrimagnetic state, in which the V $3d$ spins align antiferromagnetically with the Mn $3d$ spins, with small canting arising from spin--orbit coupling [Fig.~\ref{fig:DFT}(a)]. In this case, the Hubbard $U$ effectively enhances the crystal-field splitting associated with the $c$-axis compression, leading to orbital ordering consisting of $\psi_1$ and $\psi_2$ [Fig.~\ref{fig:DFT}(b) and (c)]. The crucial role of the $c$-axis compression becomes clearer when the orbital fillings are viewed in the original $t_{2g}$ basis: the filling of the $d_{xy}$ orbital is close to 1, whereas those of the $d_{yz}$ and $d_{zx}$ orbitals are approximately 0.5. The resulting VED is shown in Fig.~\ref{fig:DFT}(d); however, it is inconsistent with the experimental observation.\\
\hspace*{1em} The other solution is a noncoplanar ferrimagnetic state, in which the V magnetic moments exhibit a two-in--two-out pattern [Fig.~\ref{fig:DFT}(e)]. In this case, $\psi_1$ and $(\psi_2 + i\psi_3)/\sqrt{2}$ constitute the orbital order [Fig.~\ref{fig:DFT}(f) and (g)]. The complex orbital $(\psi_2 + i\psi_3)/\sqrt{2}$ carries orbital angular moment oriented toward the $\langle 111\rangle$ direction. This direction is opposite to that of the two-in--two-out spin moment. Therefore, spin--orbit coupling is the key interaction that triggers the orbital ordering, assisted by the gap enhancement due to the Hubbard $U$. Figure~\ref{fig:DFT}(h) shows the calculated VED, which is in good agreement with the experimental result.

\clearpage

\clearpage

\begin{figure}[htbp]
\centering
\includegraphics[width=\columnwidth]{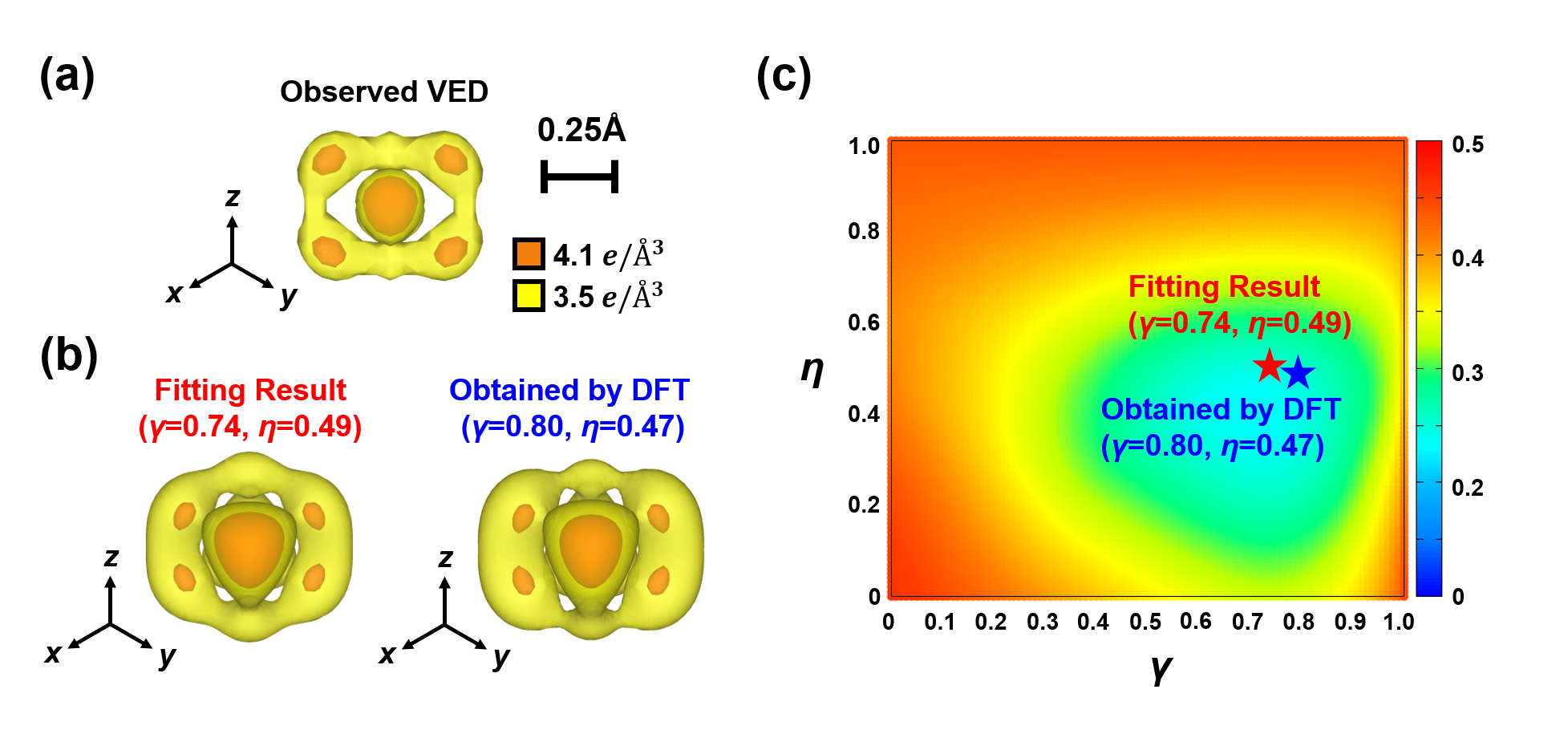}
\caption{(a) Experimentally obtained VED distribution around the V site in the Tetragonal phase. (b) Simulation of the VED distribution using parameters obtained from fitting and DFT calculations. The iso-density surface at 3.5 e/\AA$^3$ and 4.1 e/\AA$^3$ are displayed in yellow and orange, respectively. (c) Colorplot representing the dependence of evaluation function s on the parameters $\gamma$ and $\eta$ in Eqs.(1) and (3). The red and blue stars indicate the optimized parameters obtained from the fitting and the DFT calculations, respectively.}
\label{fig:fitting_vs_DFT}
\end{figure}

\hspace*{1em} Figure~\ref{fig:fitting_vs_DFT}(b) also shows simulated VED distribution using parameters obtained from fitting and DFT calculations. Both of these models accurately reproduce the experimentally obtained VED distribution anisotropy [Fig.~\ref{fig:fitting_vs_DFT}(a)]. Figure~\ref{fig:fitting_vs_DFT}(c) demonstrates that the quantum parameters obtained from experimental measurements and those calculated via DFT are in good agreement across the parameter space.

